\documentclass[aps,onecolumn,preprintnumbers,showpacs,showkeys,nofootinbib,
superscriptaddress  
]{revtex4}

\usepackage{epsfig}
\usepackage{amssymb,amsmath,amsfonts,amsthm,graphicx,psfrag}
\newcommand{\beq}{\begin{eqnarray}}
\newcommand{\eeq}{\end{eqnarray}}

\begin{document}
\preprint{}

\title{ Catalysis of Dynamical Chiral Symmetry Breaking by Chiral Chemical Potential. }

\author{V.~V.~Braguta}
\email[]{braguta@itep.ru}
\affiliation{Institute for High Energy Physics NRC “Kurchatov Institute”, 142281 Protvino, Russia}
\affiliation{Institute of Theoretical and Experimental Physics, 117259 Moscow, Russia}
\affiliation{Far Eastern Federal University,  School of Natural Sciences, 690950 Vladivostok, Russia} 
\affiliation{Moscow Institute of Physics and Technology, Institutskii per. 9, Dolgoprudny, Moscow Region, 141700 Russia}

\author{A.~Yu.~Kotov}
\email[]{kotov@itep.ru}
\affiliation{Institute of Theoretical and Experimental Physics, 117259 Moscow, Russia}
\affiliation{National Research Nuclear University MEPhI (Moscow Engineering Physics Institute), Kashirskoe Highway, 31, Moscow 115409, Russia}

\begin{abstract}
In this paper we study the properties of media with chiral imbalance
parameterized by chiral chemical potential. It is shown that depending on the strength of interaction between constituents 
in the media the chiral chemical potential
either creates or enhances dynamical chiral symmetry breaking. 
Thus the chiral chemical potential plays a role of the catalyst of dynamical chiral symmetry breaking.
Physically this effect results from the appearance of the Fermi surface and additional fermion states on this surface which take part 
in dynamical chiral symmetry breaking. An interesting conclusion which can be drawn is that  at sufficiently small temperature
chiral plasma is unstable with respect to condensation of Cooper pairs and dynamical chiral symmetry breaking
even for vanishingly small interactions between constituents.
\end{abstract}



\maketitle

{\bf Introduction.} Properties of media with nonzero chirality attract considerable interest. 
This interest is caused by unusual phenomena which take place in such media. The most renowned example 
of such phenomena is chiral magnetic effect (CME)\cite{Fukushima:2008xe, Vilenkin:1980fu}, 
which consists in the appearance of electric current in chiral medium
along applied magnetic field. The other examples of phenomena, which take place in chiral media,
are chiral vortical effect\cite{Son:2009tf, Vilenkin:1979ui, Banerjee:2008th, Landsteiner:2011cp},
chiral separation effect \cite{Son:2004tq, Metlitski:2005pr}, different chiral waves\cite{Kharzeev:2010gd, Chernodub:2015gxa, Yamamoto:2015ria} 
and others \cite{Rajagopal:2015roa,Sadofyev:2015hxa}.

Chiral media can be realized in heavy-ion collisions \cite{Kharzeev:2007jp}, Dirac semimetals 
\cite{Li:2014bha},  
Weyl semimetals 
\cite{Huang}, 
in Early Universe 
\cite{Vilenkin:1982pn}, 
in neutron stars and supernovae \cite{Charbonneau:2009ax, Ohnishi:2014uea}. 
So, chiral medium can be realized in various physical systems and it is important to study its properties.

Chiral imbalance in medium can be controlled by the chiral chemical potential -- $\mu_5$. 
The influence of the chiral chemical potential on the properties of QCD-like models was studied within lattice simulation 
\cite{Braguta:2015sqa, Braguta:2015zta, Braguta}, Dyson-Schwinger equations \cite{Wang:2015tia, Xu:2015vna}, 
different effective models \cite{Fukushima:2010fe, Chernodub:2011fr, Gatto:2011wc, Andrianov:2012dj, Andrianov:2013dta, Chao:2013qpa, Yu:2015hym}
and  universality of phase diagrams in QCD and QCD--like theories through the large--$N_c$ equivalence \cite{Hanada:2011jb}.
The results obtained within different approaches contradict to each other. For instance, some works predict 
enhancement and some suppression of the chiral symmetry breaking due to $\mu_5\neq0$. 

In this paper we are going to study the properties of the chirally imbalanced systems (not only QCD-like models), in particular, weakly interacting 
and strongly interacting chiral media. Chiral imbalance is introduced into the system by nonzero chiral chemical potential. For the systems under consideration we show that the chiral chemical potential leads either to creation or enhancement of dynamical chiral symmetry breaking. In other words, the chiral chemical potential acts as the catalyst of dynamical chiral symmetry breaking. The physical mechanism of this phenomenon is the emergence of the Fermi surface and appearance of additional fermion states on this surface which take part in dynamical chiral symmetry breaking.

{\bf NJL model and gap equation.}
To carry out our study we are going to use NJL model with the $U_L(1)\times U_R(1)$ chiral symmetry in the presence of the chiral chemical potential $\mu_5$. 
Although we use the NJL model below it will be shown that our main results are model independent. 
The Euclidean action for the NJL model can be written as 
\beq
S_E=\int d^4 x \biggl ( \bar \psi \bigl (\hat \partial   - \mu_5 \gamma_4 \gamma_5 ) \psi - G \bigr [ (\bar \psi \psi)^2 + (\bar \psi i \gamma_5 \psi)^2 \bigl ] 
\biggr ).
\label{njl}
\eeq
In this paper we will assume that the fermion fields have color indexes $\alpha=1,..,N_c$ and these indexes are contracted. 
To get rid of the four-fermion interactions one introduces scalar and pseudoscalar fields $\sigma, \pi$ which
allows to rewrite the Euclidean action  (\ref{njl}) in the following form
\beq
S_E=\int d^4 x \biggl ( \bar \psi \bigl ( \hat \partial   - \mu_5 \gamma_4 \gamma_5  +  \sigma + i \gamma_5 \pi  \bigr ) \psi + \frac 1 {4G} 
\bigr [ \sigma^2 + \pi^2 \bigl ] \biggr ).
\label{njl1}
\eeq
Integrating out fermions it is easy to get effective action $S_{eff}$ for the fields $\sigma, \pi$ 
\beq
S_{eff} &=& \int d^4 x \biggl ( \frac 1 {4G} (\sigma^2 + \pi^2) - \mbox{Tr} \log \bigr ( \hat \partial - \mu_5 \gamma_4 \gamma_5  +  \sigma + i \gamma_5 \pi  \bigl ) \biggr ) \nonumber \\ 
&=& \int d^4 x \biggl ( \frac 1 {4G} (\sigma^2 + \pi^2) - N_c \int \frac {d^4 k} {(2 \pi)^4} \sum_{s=\pm 1} \log \bigr (k_4^2 +(|k|-s \mu_5)^2 +\sigma^2 + \pi^2\bigl ) \biggr ).
\eeq
It is seen that effective action depends on $\sigma^2 + \pi^2$. Thus one can choose any direction of the condensate in $(\sigma, \pi)$--plane. 
Below we choose $\sigma$ direction of condensation leaving $\pi=0$. For zero bare fermion mass the value of $\sigma$ equals to the value of 
dynamical mass $M=\sigma$.

In the limit $N_c \to \infty$ the path integral over the fields $\sigma$ is dominated by the stationary point ${\delta S_{eff}} / {\delta \sigma}=0$.
The value of dynamical fermion mass $M$ can be determined from the gap equation
\beq
 \frac {1} {G N_c} = \frac 1 {\pi^2} \int_0^{\Lambda} k^2 dk \biggr [ \frac 1 {\sqrt{(|\vec k|-\mu_5)^2+M^2}} + \frac 1 {\sqrt{(|\vec k|+\mu_5)^2+M^2}} \biggl ]
\label{sigma1}
\eeq
Notice that in last equation we used three-momentum cutoff regularization scheme.
Assuming that $M, \mu_5 \ll \Lambda$ equation (\ref{sigma1}) can be written as 
\beq
\frac {1} {\alpha_{NJL}} - 1 = \biggl ( y^2- \frac {x^2} 2 \biggr ) \log {\frac 1 {x^2}} 
\label{gap}
\\ \nonumber
\alpha_{NJL} = \frac {G N_c \Lambda^2} {\pi^2},~~x =\frac {M} {\Lambda},~~y =\frac {\mu_5} {\Lambda}
\eeq
At zero chiral chemical potential equation (\ref{gap}) coincides with the NJL gap equation 
for dynamical fermion mass ( see review 
\cite{Klevansky:1992qe}). 
For $\mu_5=0$ and $\alpha_{NJL}<1$ left side of equation (\ref{gap}) is positive, 
but right side is negative. So, equation (\ref{gap}) has the solution $M\neq0$ only for sufficiently 
strong interaction $\alpha_{NJL}>1$. Below it will be shown that nonzero $\mu_5$ changes the properties 
of the gap equation (\ref{gap}) dramatically. 

{\bf Weakly interacting chiral medium.} First we are going to consider
the case of weak interaction $\alpha_{NJL} \ll 1$. There is no solution of the gap equation (\ref{gap})
for $\mu_5=0$, i.e. dynamical mass is zero $M=0$. However, for any $\mu_5 \neq 0$ and $\alpha_{NJL} \ll 1$ 
there appears a solution which can be written as 
\beq
M^2 = \Lambda^2 \exp {\biggl [ - \frac {\pi^2} {G N_c \mu_5^2} \biggr ]}.
\label{mdyn}
\eeq
Last equation shows that nonzero chiral chemical potential leads to dynamical chiral symmetry breaking and generation of fermion mass even 
for vanishing attraction between fermions.
The dynamical mass $M$ for weak interaction $\alpha_{NJL}\ll1$ is determined 
by the behavior of the system near the Fermi surface $|\vec k|=\mu_5$. To see this we write 
the gap equation (\ref{sigma1}) as
\beq
 \frac {1} {G N_c} \approx \frac 1 {\pi^2} \int_0^{\Lambda} k^2 dk  \frac 1 {\sqrt{(| k|-\mu_5)^2+M^2}}  \approx
\frac {\mu_5^2} {\pi^2} \int_{\mu_5-\delta}^{\mu_5+\delta}  dk  \frac 1 {\sqrt{(|k|-\mu_5)^2+M^2}} \approx \nu(E_F) \log {(M^2)},
\label{sigma2}
\eeq
where $\nu({E_F})$ is a density of states on the Fermi surface
\beq
\nu({E_F}) = \frac 1 V \frac {d N(E)} {d E} \biggr |_{|\vec k|=\mu_5} = \frac {\mu_5^2} {\pi^2}.
\label{nu}
\eeq
Expression (\ref{sigma2}) represents BCS instability on the Fermi surface. Dynamical mass (\ref{mdyn}) written 
in terms of the $\nu(E_F)$ is $M^2 = \Lambda^2 \exp { ( - 1/  {G N_c \nu(E_F)} )}$, what is very similar 
to the mass gap in the BCS theory of superconductivity $\Delta = \omega_D \exp { ( - const/  {G_S  \nu_F} )}$\cite{Bardeen:1957mv},
where $\omega_D$ is the Debye frequency, $G_S$ is a coupling constant and $\nu_F$ is the density of states on the Fermi
surface.

{\bf Strongly interacting chiral medium.} We proceed with the study of the solution of the gap equation (\ref{gap})
for strong interaction $\alpha_{NJL} \sim 1$. First let us consider the case when the interaction is strong 
but insufficient for dynamical chiral symmetry breaking without the chiral chemical potential:~ $0<1-\alpha_{NJL} \ll 1$. 
Similarly to weakly interacting medium $\mu_5 \neq 0$ leads to chiral symmetry breaking and 
generation of fermion mass which can be obtained from equation (\ref{gap})
\beq
M^2 \simeq 2 \mu_5^2 
\label{mass1}
\eeq
Notice that the correction to equation (\ref{mass1}) is $\sim 1/N_c$. 
Since our consideration is valid only at the leading order approximation in $1/N_c$ expansion,
we don't show it here.

Now let us consider the case $\alpha_{NJL}>1$. We have already mentioned above that if $\alpha_{NJL} > 1$ there is 
a solution of the gap equation (\ref{gap}) for zero chiral chemical potential which will be designated as $M_0$. 
We start our study from the case of small chiral chemical potential $\mu_5 \ll M_0$. Expanding equation (\ref{gap}) in 
the vicinity of the solution $M_0$ one gets
\beq
M^2 \simeq M_0^2  + 2  {\mu_5^2}.
\label{mass2}
\eeq
So, one sees that $M$ is quadratically rising function of the $\mu_5$ at small $\mu_5$.
In the case of large chemical potential $\mu_5 \gg M_0$ dynamical fermion mass is given by formula (\ref{mass1}).

Notice that in the derivation of formulas (\ref{mass1}) and (\ref{mass2}) the term which represents logarithmic singularity
(the term $\sim y^2$ in equation (\ref{gap})), i.e. dynamics near the Fermi surface,  plays a crucial role.

{\bf Chiral medium and BCS theory of superconductivity.} The BCS instability and the formula for the dynamical mass (\ref{mdyn})
allow us to state that there is an analogy between NJL model with the chiral chemical potential 
and BCS theory of superconductivity. To see this explicitly we are going to use variational approach. 
First note that in the NJL model without interaction $G=0$, at $T=0$ and $\mu_5>0$ vacuum state -- $| p_F \rangle$
is two Fermi spheres of right particles and right antiparticles with radius $\mu_5$. It is reasonable to assume that interacting 
NJL supports condensation in the right quark--right antiquark channel. A suitable vacuum state for this model can be written as
\beq
\label{vacuum}
| vac \rangle &=& \hat G_1 \hat G_2 \hat G_3| p_F \rangle, \\ \nonumber
\hat G_1 &=& \prod_p \biggl ( \cos {(\theta_L )} - \sin {(\theta_L )} {\hat a}^+_{L,p} {\hat b}^+_{L,-p} \biggr ), \\ \nonumber
\hat G_2 &=& \prod_{p>\mu_5} \biggl ( \cos {(\theta_R )} + \sin {(\theta_R )} {\hat a}^+_{R,p} {\hat b}^+_{R,-p} \biggr ), \\ \nonumber
\hat G_3 &=& \prod_{p<\mu_5} \biggl ( \cos {(\tilde \theta_R )} + \sin {(\tilde \theta_R )} {\hat b}_{R,-p} {\hat a}_{R,p} \biggr ),
\eeq
where $({\hat a}^+_{L,p},{\hat a}_{L,p})$/ $({\hat b}^+_{L,p},{\hat b}_{L,p})$ are creation, 
annihilation operators for left particles and antiparticles, $({\hat a}^+_{R,p},{\hat a}_{R,p})$/ $({\hat b}^+_{R,p},{\hat b}_{R,p})$ are creation, 
annihilation operators for right particles and antiparticles correspondingly\footnote{Note that in the variation approach in addition to the parameters 
$\theta_L, \theta_R, \tilde \theta_R$ one introduces relative phases between $\cos{(\theta)}, \sin{(\theta)}$ terms. We have checked 
that trial vacuum state (\ref{vacuum}) gives minimum energy with respect to the variation over these phases.}. 
Note that the operator $\hat G_1$ creates left states, the operator $\hat G_2$ creates right states above the Fermi surface
and the operator $\hat G_3$ creates right states below the Fermi surface. The vacuum energy of NJL model (\ref{njl})
can written as 
\beq
E_{vac} &=& 2N_c \int_{p<{\mu_5}} \frac {d^3 p} {(2 \pi)^3} (p-\mu_5) \cos^2 {\tilde \theta_R} + 
2N_c \int_{p>{\mu_5}} \frac {d^3 p} {(2 \pi)^3} (p-\mu_5) \sin^2 {\theta_R} +
2N_c \int \frac {d^3 p} {(2 \pi)^3} (p+\mu_5) \sin^2 {\theta_L}  \nonumber \\
&-& G N_c^2 \biggr (  \int_{p<{\mu_5}} \frac {d^3 p} {(2 \pi)^3} \sin {2 \tilde \theta_R} + \int_{p>{\mu_5}} \frac {d^3 p} {(2 \pi)^3} \sin {2  \theta_R} +
\int \frac {d^3 p} {(2 \pi)^3} \sin {2 \theta_L} \biggl )^2
\eeq
The variation of the $E_{vac}$ with respect to the parameters $\theta_L, \theta_R, \tilde \theta_R$ gives the following 
values
\beq
\label{theta}
\tan {2 \theta_L} = 2 G N_c \frac {\Delta} {p+\mu_5},~~~ \tan {2 \theta_R} = 2 G N_c \frac {\Delta} {p-\mu_5},~~~ \tan {2\tilde \theta_R} = 2 G N_c \frac {\Delta} {\mu_5-p}, \\
\Delta= \biggr (  \int_{p<{\mu_5}} \frac {d^3 p} {(2 \pi)^3} \sin {2 \tilde \theta_R} + \int_{p>{\mu_5}} \frac {d^3 p} {(2 \pi)^3} \sin {2  \theta_R} +
\int \frac {d^3 p} {(2 \pi)^3} \sin {2 \theta_L} \biggl ).
\label{Mmass}
\eeq
Substituting the values of the $\theta_L, \theta_R, \tilde \theta_R$ from (\ref{theta}) to expression (\ref{Mmass}) 
one finds gap equation (\ref{gap}) and dynamical fermion mass $M=2 G N_c \Delta$.

From this consideration one sees that the energy minimum of the NJL model with $\mu_5 > 0$ is realized through 
the condensation of the Cooper pairs which consist of right particle and right antiparticle and break chiral symmetry. 
The energy in the minimum is smaller than the energy of free chiral plasma. 
Thus chiral plasma is unstable with respect to chiral symmetry breaking and condensation of the Cooper pairs.

Note also that for systems considered in this paper the chiral chemical potential dynamically breaks chiral symmetry, if it was not broken, or strengthens it otherwise. Physically this effect stems from the formation of the Fermi surface and appearance of additional fermion states on
this surface which take part in dynamical chiral symmetry breaking. For this reason the chiral chemical potential plays 
a role of the catalyst of dynamical chiral symmetry breaking.

 In the above considerations we used NJL model. However, it is clear that in any model nonzero $\mu_5$  
leads to Fermi surface with additional fermion states which due to BCS instability create or enhance chiral symmetry breaking. 
We believe that this effect is universal, i.e., model independent.

{\bf Confirmation from lattice and comparison with other studies.} NJL model in the strong coupling regime is believed to be an effective low energy model of QCD. 
So, it is interesting to compare the predictions of the NJL model with the results of lattice study 
of QCD-like models \cite{Braguta:2015sqa, Braguta:2015zta, Braguta}. 

\begin{figure}[t]
        \includegraphics[scale=0.4, angle=270]{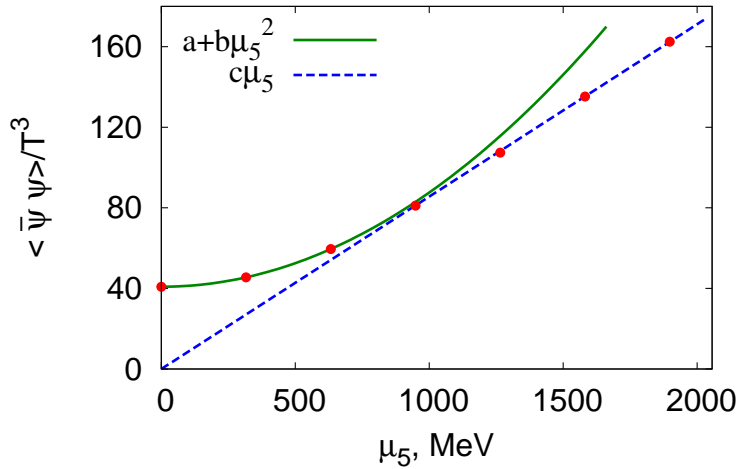}
        \caption{Chiral condensate as a function of $\mu_5$ in the confinement phase ($T=158$ MeV).}
        \label{fig:1}
\end{figure}

One of the prediction of NJL model is that dynamical fermion mass  quadratically rises at small $\mu_5$ 
which then switches to the linear rising behavior at large $\mu_5$. This is very important property
since at small $\mu_5$ the vacuum rearrangement is due to the strong interaction and at
large $\mu_5$ the vacuum rearrangement is due additional fermion states on the Fermi surface. 
In the NJL model $M \sim \langle \bar \psi \psi \rangle$, 
so the behavior of the dynamical mass and chiral condensate must be similar. In Fig.\ref{fig:1} we plot 
$\langle \bar \psi \psi \rangle$ as a function of $\mu_5$ in the confinement phase 
obtained in paper\cite{Braguta:2015zta}. It is seen that the points with $\mu_5<1000$ MeV lie on the curve 
$a+b \mu_5^2$ and the points $\mu_5>1000$ GeV lie on the curve $c \mu_5$. 
The statistical uncertainties in the data are $\sim 0.1 \%$. 
So lattice data confirms the results of this paper.

Further we notice that the larger $\mu_5$ the larger dynamical fermion mass $M$. 
This allows us to expect the rising of the critical temperature of breaking/restoration 
of chiral symmetry as the chiral chemical potential grows. The property is in agreement with the results of 
papers \cite{Braguta:2015sqa, Braguta:2015zta, Braguta}. 

We have already mentioned that there are a lot of papers where the influence of the chiral chemical potential on 
the properties of QCD-like models was studied.  In addition to lattice papers \cite{Braguta:2015sqa, Braguta:2015zta, Braguta}, 
there are studies based on Dyson-Schwinger equations \cite{Wang:2015tia, Xu:2015vna}, 
the large--$N_c$ equivalence between QCD--like theories\cite{Hanada:2011jb}.
The results of these papers support our conclusions. 

The results obtained within different effective models \cite{Fukushima:2010fe, Chernodub:2011fr, Gatto:2011wc, 
Andrianov:2012dj, Andrianov:2013dta, Chao:2013qpa, Yu:2015hym} strongly depend 
on the details of these models. The authors of paper \cite{Yu:2015hym} studied how the results depend on 
regularization in NJL model. In particular, they showed that chiral condensate is rising function of $\mu_5$ for 
$\mu_5<\mu_5^c \sim \Lambda$ and decreasing function of $\mu_5$ for $\mu_5>\mu_5^c$. 
We believe that the disagreement between our results and the results of paper \cite{Yu:2015hym} can be explained as 
follows. In our paper we have shown that dynamics near the Fermi surface $|\vec k|=\mu_5$ is very important for correct description 
of chiral media.
If $\mu_5\sim \mu_5^c \sim \Lambda$, the ultraviolet cutoff $\Lambda$ effectively cuts 
important degrees of freedom near the Fermi surface leading to incorrect result. For this reason 
if $\mu_5 \ll \Lambda$ our result is in agreement with \cite{Yu:2015hym}. So, our
conclusion is that to get dependable results within effective models all energy scales 
in these models must be much smaller than the ultraviolet cutoff $\Lambda$. 

{\bf Conclusion.} In this paper we have studied the properties of the media with chiral imbalance 
parameterized by the chiral chemical potential. We considered two cases: weakly interacting 
and strongly interacting chiral media. For all considered systems we have shown 
that the chiral chemical potential either creates or enhances 
dynamical chiral symmetry breaking. The mechanism responsible for this phenomenon 
is the appearance of the Fermi surface and
additional fermion states on this surface which take part in dynamical chiral symmetry breaking. 
So, the chiral chemical potential plays a role of the catalyst of dynamical chiral 
symmetry breaking.

An interesting observation which follows from the results of this paper is that 
at sufficiently small temperature weakly interacting chiral plasma is unstable 
with respect to condensation of Cooper pairs, dynamical chiral symmetry breaking and generation of 
dynamical fermion mass. We believe that appearance of chiral symmetry breaking and generation of 
fermion mass in chiral plasma may considerably modify CME and other phenomena which take place 
in chiral media.

\section*{Acknowledgments}

The authors are grateful to V. I. Zakharov and M. N. Chernodub for interesting and stimulating discussions. 
The work was supported by Far Eastern Federal University, Dynasty foundation, 
by RFBR grants 15-02-07596, 15-32-21117, 16-32-00048 and by a grant of the FAIR-Russia Research Center.

\end{document}